
\magnification=1000
\parindent=0pt
\overfullrule=0pt
\def\square{\kern1pt\vbox
            {\hrule height 0.6pt\hbox{\vrule width 0.6pt\hskip 3pt
 \vbox{\vskip 6pt}\hskip 3pt\vrule width 0.6pt}\hrule height 0.6pt}\kern1pt}
\def\dt{\hbox{det}}
\def\Bt{B\"acklund transformation }
\def\Bts{B\"acklund transformations }
\def\half{{1\over2}}

\def\p{\partial} \def\D{{\cal D}}  \def\cD{{\bar {\cal D}}}
\def\der#1{{\partial \over \partial #1}}
\def\dd#1#2{{\partial #1 \over \partial #2}}
\def\bD{{\bar D}}
\def\bt{{\bar \theta}}
\def\t{\theta}	   \def\e{\epsilon}
\def\g{\gamma} \def\b{\beta} \def\a{\alpha} \def\l{\lambda}
\def\d{\delta} 	   \def\s{\sigma}   \def\r{\rho}
\def\P{\varphi} \def\m{\mu} \def\n{\nu}  \def\da{{\dot \alpha}}
\def\db{{\dot \beta} } \def\2{{\dot 2}} \def\1{{\dot 1}}
\rightline{hep-th/9301098}
\rightline{IHEP 92-170}
\vskip 1.5 truecm
\centerline{{\bf
B\"acklund transformation for supersymmetric self-dual theories }}
\centerline{{\bf
for semisimple gauge groups and a hierarchy of $A_1$ solutions  }}
\vskip 1.5 truecm
\centerline{Ch. DEVCHAND$^{1,}$\footnote{*}
{electronic address: devchand@theor.jinrc.dubna.su}\footnote{}{
postal address: LTPh, JINR, P.O. Box 79, Head Post Office, 101 000 Moscow,
Russia.}
            and A.N. LEZNOV$^{2}$}
\vskip .5 truecm
\centerline{{\it  $^{1}$Laboratory of Theoretical Physics,  }}
\centerline{{\it   Joint Institute for Nuclear Research,      }}
\centerline{{\it          Dubna, 141980 Russia.            }}
\vskip .35 truecm
\centerline{{\it  $^{2}$Institute for High Energy Physics,  }}
\centerline{{\it                 Protvino, 142284   Russia.   }}
\vskip 3 truecm
\leftline{{\bf Abstract}}
We present a B\"acklund transformation (a discrete symmetry transformation)
for the self-duality equations for supersymmetric gauge theories in
N-extended super-Minkowski space ${\cal M}^{4|4N}$ for an arbitrary
semisimple gauge group. For the case of an $A_1$ gauge algebra we integrate
the transformation starting with a given solution and iterating the process
we construct a hierarchy of explicit solutions.
\vskip 20pt
{\bf 1. Introduction}
\vskip 20pt
The self-duality equations for supersymmetric gauge fields in superspace
have a construction of solutions very similar to that for non-supersymmetric
gauge fields. This is not surprising since, once the kinematic constraints
have been solved, the dynamical equations for the superfield ``prepotential''
describing the theory have a form independent of the extension of
superspace, holding for the non-supersymmetric case as well. The discrete
symmetry transformation (B\"acklund transformation)
for non-supersymmetric self-dual gauge fields taking values in
an arbitrary semisimple Lie algebra recently presented [6] therefore
actually holds, with appropriate replacement of ordinary fields by
superfields, for the supersymmetric cases as well. The transformation,
however, may be written in a simpler (lower
order) form allowing more compact proofs and yielding the previous results
for the non-supersymmetric (N = 0) self-duality equations as direct
corollaries.
We present the B\"acklund transformation for an arbitrary
semisimple gauge group in section 3 and present a simple expression for the
change induced by the transformation in the topological charge density.
Specialising in section 4 to an $A_1$ gauge theory, we write an explicit
form of the transformation for the gauge algebra components of the superfield
prepotential.  Starting from a given solution for the restriction of
the gauge algebra to a solvable sub-algebra of $A_1$, the transformation is
integrable, yielding a new solution  which spans the entire $A_1$ algebra. In
fact, the B\"acklund transformation applied to the thus obtained solution can
also be integrated; and this process can be iterated to
construct a hierarchy of explicit solutions, which we present in section 5.
Even for the N=0 case our solutions generalise those obtained in [2]
from an explicit construction of the Atiyah-Ward ans\"atze.
\vskip 20pt {\bf 2. The dynamical self-duality equations}
\vskip 20pt
In N-extended complexified super-Minkowski space ${\cal M}^{4|4N}$ of
complex dimension $(4| 4N)$ with coordinates
$\{x^{\a\db}, \t^\a_s, \bt^{\da t}\}$, where
$\a,\da$ are two-component spinor
indices and $s,t = 1,....,N$ are internal indices, the upper and lower ones
referring to fundamental and conjugate representations respectively,
the self-duality equations
may be written in the form of the following
super-curvature constraints in N-extended superspace (e.g.[10]):
$$\eqalign{
\{ \D_{\a s} ~,~\D_{\b t}\} =&\ 0
\cr
\{ \cD^{(s}_{\da} ~,~ \cD^{t)}_{\db} \} =&\ 0
\cr
\{ \D_{\a s} ~,~ \cD^t_{\db} \} =&\  2 \delta^t_s \D_{\a \db}
\cr
 [\D_{\b s} ~,~\D_{\a \da}] =&\ 0
,\cr} \eqno(1)$$
where $\D_A \equiv \p_A + A_A =
\{\D_{\a\db} \equiv \s^\mu_{\a\db}\D_\mu , \D_{\a s}, \cD^t_{\db}\} $
are gauge-covariant derivatives and the supertranslation vector fields
$$\p_A  =\{\p_{\a\db}, D_{\a s}, \bD^t_\db \}
\equiv \{\der {x^{\a\db}} ~,~ \der {\t^{\a s}} ~,~
\der {\bt^{\db}_t} + 2 \t^{\a t} \p_{\a\db}, \} $$ realise the superalgebra
$$\{D_{\a s} ~,~D_{\b t}\} =\ 0
\ =\  \{ \bD^s_{\da} ~,~\bD^t_{\db}\} $$
$$\{D_{\a s} ~,~\bD^t_{\db} \} =\  2\  \d^t_s \p_{\a\db}  $$
$$[D_{\b s} ~,~\p_{\a\db}] =\ 0 =\  [\bar D^t_{\da} ~,~\p_{\a\db}] . $$
Using Jacobi identities, these equations for the superconnection $A_A$
may easily be seen to imply the familiar self-duality relations
$$ F_{\mu \nu } = \half \e_{\mu \nu \rho \s} F_{\rho \s}     $$
for the field strengths.
A large number of the superconnection components are clearly pure-gauges,
containing no dynamical information.
In particular, in the supersymmetric form of the Yang gauge [13]
$$ A^s_\a = 0 = A_{\1 t} = A_{\a \1} ,\eqno(2)$$
which is consistent with (1), the dynamical content of (1)
is contained in the supersymmetric Yang equations [3]
$$ A_{\2 s} = g^{-1} \bD_{\2 s} g  = \bD_{\1 s} f    ,\eqno(3a)$$
where $g$ is a matrix superfield in the gauge group and  $f$ takes values
in the gauge algebra.
Consistency conditions for these equations include all but
the following residual constraints from (1):
$$
F^{s}_{\a{\dot 2} t}~=~0  \hbox { i.e. }
A_{\a{\dot 2}} ~=~ {1\over 2N}\delta^s_t  D_{\a s} D^t_{{\dot 1}} f
\eqno(4)$$
and $$
F^s_{\a \b{\dot 2}}~=~ D^s_{\a} A_{\b{\dot 2}} ~=~ 0
.\eqno(5)$$
Now if $f$ is taken to be a chiral superfield satisfying $D_{\a s}f = 0, $
(4) yields vector potentials
$$
A_{\a{\dot 2}} ~=~g^{-1} \p_{\a\2} g  =  \p_{\a{\dot 1}} f
\eqno(3b)$$
which are automatically chiral i.e. satisfying (5). These are precisely the
usual ($N=0$) Yang equations [1,9].
The only conditions for $f$ which are not automatic are those arising from
the Maurer-Cartan identities for $g$, viz.,
$$ D^{(s}_{{\dot 2}} D^{t)}_{{\dot 1}} f ~
+~D^{(s}_{{\dot 1}} f ~D^{t)}_{{\dot 1}} f ~~=~0 \eqno(6a)$$
from (3a) and
$$
\partial_{\a \db} \partial^{\a\db}f
= [ \p_{\a{\dot 1}}f~,~ \p^\a_{\dot 1}f]
\eqno(6b)$$ from (3b);
the latter in Yang coordinates
$$ x_{\a{\dot \b}} = \pmatrix{ y&-{\bar z}\cr
                               z& {\bar y}\cr}\      $$
taking the familiar form
$$\square f \equiv (\p_y\p_{\bar y} + \p_z\p_{\bar z}) f
= [\p_y f ~,~ \p_z f] .$$
In fact, acting on (6a) by $D_{\a s} D^\a_{t}\equiv D_{1 (s} D_{2 t)}$
yields (6b), so we may
think of the former as the equations of motion for the N = 2,3 and 4
theories; and the latter as the equations of motion for the N = 0 and 1
cases. These equations are Euler-Lagrange equations for the functional [7]
$$\eqalign{
{\cal L} =&  tr {1\over 2 N(N+1)} D_1^{(s} D_2^{t)}
        ({1\over 2} D_{{\dot 1} s}f D_{{\dot 2} t}f +
        {1\over 3} f D_{{\dot 1} s}f D_{{\dot 1} t}f ) \cr
  =& tr  \half     (\partial_{\a \1}f \partial^{\a}_{\2}f +
        {1\over 3} f  \partial_{\a\1}f  \partial^{\a}_\1 f )
,\cr}\eqno(7)$$
the second expression following on performing the two odd differentiations
for arbitrary N and using the chirality of $f$.

It is results for the dynamical form (6b) of the Yang equations advertised
in [6] which we presently supersymmetrise. We shall present explicit proofs
of the symmetry transformations mainly for the odd equation (6a), the
statements in [6] for equation (6b) clearly following as corollaries.
\vskip 20pt
{\bf 3. B\"acklund transformation for an arbitrary semisimple gauge
group}
\vskip 20pt
We consider the system of equations
$$
 D^{(s}_{{\dot 2}} D^{t)}_{{\dot 1}} f ~
+~\{ D^{s}_{{\dot 1}} f~,~ D^{t}_{{\dot 1}} f \}~~=~0 \ ;\ D_{\a s}f ~=~0
,\eqno(8)$$
where $f$ takes values in the algebra of a (semi-simple) gauge group.
An auto-b\"acklund transformation for these equations
is given by the following system of equations for $F$ and a
matrix function $S$ in the gauge group:
$$ D_\1^s F \ =\ S D^s_\1 \tilde f S^{-1} \ -\  D_\2^s S~ S^{-1} \ ;\
   D_{\a s}F ~=~0 ,\eqno(9a)$$
where $\tilde f = r f r^{-1}$, the conjugation denoting an automorphism
of the gauge algebra, which we shall take to be one which maps the maximal
root to the minimal one.
The proof that $F$ satisfies (8) if $f$ does is straightforward;
consistency of (9a) requiring that
$$
 D^{(s}_{{\dot 2}} D^{t)}_{{\dot 1}} F ~
+~\{ D^{s}_{{\dot 1}} F~,~ D^{t}_{{\dot 1}} F \}~~=~
S r~ ( D^{(s}_{{\dot 2}} D^{t)}_{{\dot 1}} f ~
  +~\{ D^{s}_{{\dot 1}} f~,~ D^{t}_{{\dot 1}} f \} ) ~r^{-1}S^{-1}
.$$
Integrability of (9a) also requires that S satisfies the equation
$$
D^{(s}_\2(S^{-1} D^{t)}_\1 S) \
+\ \{ S^{-1} D^{(t}_\1 S \ ,\ D^{s)}_\1 \tilde f\} = 0 .\eqno(10a)$$
In order to have an explicit B\"acklund transformation, we need to insert a
solution of this equation into (9a) . An ansatz solving (10a) is given by
$$  S^{-1} D^{t}_\1 S\ =\  {1\over f^+} [X_M^+ , D^t_\1 \tilde f ] \
                           -\ D_\2^t  \big({1\over f^+} \big) X_M^+
,\eqno(11a)$$
where $X^+_M$ is the normalised algebra element corresponding to the maximal
root and $f^+$ is its coefficient in $f$. This ansatz is consistent
in virtue of the equations of motion for $f^+$.

As a corollary, since $F,f$ and $S$ are chiral,
we clearly have that the system:
$$ \p_{\a\1} F \ =\ S \p_{\a\1} \tilde f S^{-1} \ -\  \p_{\a\2} S~ S^{-1} \
;\   \eqno(9b)$$
$$  S^{-1} \p_{\a\1} S\ =\  {1\over f^+} [X_M^+ , \p_{\a\1} \tilde f ] \
                           -\ \p_{\a\2}  \big({1\over f^+} \big) X_M^+
\eqno(11b)$$
is a B\"acklund transformation for eq.(6b); satisfying the consistency
relations
$$
\partial_{\a \db} \partial^{\a\db}F
- [ \p_{\a{\dot 1}}F~,~ \p^\a_{\dot 1}F]
= Sr (\partial_{\a \db} \partial^{\a\db}f
- [ \p_{\a{\dot 1}}f~,~ \p^\a_{\dot 1}f] ) r^{-1}S^{-1}
$$
and
$$
\p_{\a{\dot 2}} (S^{-1} \p^\a_{\dot 1} S )
{}~+~   [ S^{-1} \p_{\a\1} S \ ,\ \p^\b_{\1} f ] = 0 .\eqno(10b)$$

The topological charge density (or equivalently the Yang-Mills lagrangian
density for self-dual fields) is given by the formula [1]
$$\eqalign{
q(f) =&\  {1\over 8} tr  \e^{\m\n\r\s} F_{\m\n} F_{\r\s} \cr
     =&\  tr(F_{{\bar y}z} F_{y{\bar z}} + F_{z{\bar z}}F_{y{\bar y}} ) \cr
     =&\  tr (\p^2_z f \p^2_y f -  (\p_z\p_y f)^2  )
     ,\cr}
$$ using the Yang equations(3b).
\vskip 5pt
{\it Under the B\"acklund transformation (9,11) the topological charge
density changes according to the formula }
$$ q(F) = q(f) + \square \square ln f^+  .\eqno(12)$$
\vskip 5pt
{\it Proof.}
$$\eqalignno{
q(F)  =&\  \half tr \p_{\a\1} \p^{\g}_\1 F \p^{\a}_\1 \p_{\g_\1}F  \cr
      =&\  \half tr \p^\a_\1 \p^\g_\1  (\p_{\a\1}F \p_{\g\1}F)      \cr
 =&\  q(f) + tr  \p^\g_\1  (- \p_{\a\1}\tilde f \p^\a_\1 (S^{-1}\p_{\g\2}S))
        + \half tr \p^\a_\1 \p^\g_\1 (\p_{\a\2} S S^{-1} \p_{\g\2}S S^{-1})
,&(13) \cr} $$
using the \Bt (9b) and the cyclic property of the trace. Now using the
Maurer-Cartan identity
$$\p^\a_\1(S^{-1}\p_{\g\2}S) - \p_{\g\2}(S^{-1}\p^\a_\1 S)
+[S^{-1}\p^\a_\1 S ~,~ S^{-1}\p_{\g\2}S ]  \equiv 0  $$
the second term on the right takes the form
$$\eqalign{
tr&  \p^\g_\1  (- \p_{\a\1}\tilde f \p_{\g\2}(S^{-1}\p^\a_\1 S)
          + [\p_{\a\1}\tilde f ~,~ S^{-1}\p^\a_\1 S ] S^{-1}\p_{\g\2}S ) \cr
= tr&  \p^\g_\1  (- \p_{\a\1}\tilde f \p_{\g\2}(S^{-1}\p^\a_\1 S)
          + \p_{\a\2}(S^{-1}\p^\a_\1 S) (S^{-1}\p_{\g\2}S) )       \cr} $$
in virtue of the integrability condition (10b); and inserting
expression (11b) for $(S^{-1}\p^\a_\1 S)$ and using the property that
$tr (X^+_M \tilde f) = f^+ \equiv e^\b $, we obtain for this second term
the expression
$$ tr \p^\g_\1  ( 2 \p_{\a\2} \b \p_{\g\2}\b  e^{-\b} \p^\a_\1 e^\b
		  - e^{-\b}\p_{\g\2}\square e^\b
		  - \p_{\g\2} \p^{\a}_\2 e^{-\b} \p_{\a\1}e^\b ) .$$
Finally, the last term in (13) is equal to
$$\eqalign{
     &  \half tr \p^\a_\1 \p^\g_\1 (\p_{\a\2} S S^{-1} \p_{\g\2}S S^{-1}) \cr
     &  = \p^\a_\1 \p^\g_\1 (\p_{\a\2} \b \p_{\g\2}\b)
,\cr}$$
using (11b) and  $tr H^2 = 2$. Putting all the terms together yields the
result (12).  \quad \quad \square

\vskip 20pt
{\bf 4. Explicit B\"acklund transformation for an $A_1$ gauge theory}
\vskip 20pt
We now display the explicit form of the B\"acklund
transformation for the $A_1$ case. The equations of motion (6a) for the
coefficients of  $f$  in a Cartan-Weyl basis:
$ f \ =\ f^+X^+ + f^-X^- + f^0 H$, where $[H, X^{\pm}] = \pm 2 X^{\pm} $ and
$[X^+ , X^-] = H $ are
$$\eqalign{
 D^{(s}_{{\dot 2}} D^{t)}_{{\dot 1}} f^0 ~
+~ D^{(s}_{{\dot 1}} f^+ ~ D^{t)}_{{\dot 1}} f^- =& 0
\cr
 D^{(s}_{{\dot 2}} D^{t)}_{{\dot 1}} f^+ ~
+~2 D^{(s}_{{\dot 1}} f^0 ~ D^{t)}_{{\dot 1}} f^+ =& 0
\cr
 D^{(s}_{{\dot 2}} D^{t)}_{{\dot 1}} f^- ~
+~2 D^{(s}_{{\dot 1}} f^- ~ D^{t)}_{{\dot 1}} f^0 =& 0
,\cr}\eqno(14a)$$
where $ f^\pm $ and $ f^0 $ are chiral superfields; and equations (6b)
take the form
$$\eqalign{
\square f^0 =&\  ~~ \p_{[y} f^+  \p_{z]} f^-    \cr
\square f^+ =&\  2\  \p_{[y} f^0  \p_{z]} f^+   \cr
\square f^- =&\  2\  \p_{[y} f^-  \p_{z]} f^0    .\cr}
\eqno(14b)$$
The ansatz (11) clearly means that  $S$  takes values in the solvable part
of the algebra. Inserting the parametrisation
$ S \ =\ e^{\a_1 X^+} e^{\b H} $ in (11a)
and choosing the algebra automorphism $r$ to be such that
$$ rX^{\pm}r^{-1} = X^{\mp} ~~,~~ rHr^{-1} = -H ,$$
yields  $\b =\ \ln f^+ $  and the following equations for $\a_1$:
$$ D^t_\1 \a_1 = 2 f^+ D^t_\1 f^0 + D^t_\2 f^+  ;\eqno(15)$$
consistency clearly being guaranteed by the $f^+$ equation in (14a).

With  $\b =\ \ln f^+ $  and using the Campbell-Hausdorff relations
$$\eqalign{e^{-\b H}  X^\pm\ e^{\b H} \  =&\   e^{\mp 2\b} X^\pm \cr
            e^{\a_1 X^+} H\    e^{-\a_1 X^+} =&\  H - 2\a_1 X^+ \cr
  e^{\a_1 X^+} X^-   e^{-\a_1 X^+} =&\  X^-  + \a_1 H - 2\a_1^2 X^+ ,\cr}
\eqno(16)$$
the algebra components of (9a) may be written:
$$\eqalign{
D^s_\1 F^- =& (f^+)^{-2} D^s_\1 f^+  , \cr
D^s_\1 F^0 =& - D^s_\1 f^0 \ -\ D^s_\2 \ln f^+ + (f^+)^{-2} \a_1 D^s_\1 f^+
,\cr
D^s_\1 F^+ =& (f^+)^2 D^s_\1 f^- + D^s_\2 \a_1
            + \a_1 (D^s_\1 \big( {\a_1 \over f^+ } \big) + D^s_\2 \ln f^+)
.\cr} $$
Now the second equation above together with (15) gives the relation
$\a_1 =\  - f^+(F^0 - f^0) $, yielding the following explicit form
of the B\"acklund transformation.
\vskip 5pt
{\it If  $ \{f^0,f^+,f^-\} $  satisfy eqs.(14a,b),
then so do $\{F^0,F^+,F^-\}$ given by the following equations:}
$$\eqalignno{
 F^-  =& -{1\over f^+} &(17a)\cr
 D^s_\1 F^0 =&  - D^s_\1 f^0 \ -\ D^s_\2 \ln f^+
                  - (F^0 - f^0)D^s_\1 ln f^+  &(17b)\cr
 D^s_\1 F^+ =&  f^+ \{ (F^0 - f^0)D^t_\1(F^0 - f^0) +\  f^+  D^s_\1 f^- \
                +\  D^t_\2(F^0 - f^0) \} .&(17c)\cr} $$
\vskip 5pt

{\it Proof.} The transformation is manifestly true since it arises from
the insertion of a consistent solution of (10) in (9). The direct
verification however is not entirely staightforward.
The first equation evidently implies
$$\eqalign{
D^{(s}_\2 D^{t)}_\1 F^-
=&\ -~2 (f^+)^{-3} D^{(s}_\2 f^+ D^{t)}_\1 f^+
    + (f^+)^{-2}  D^{(s}_\2 D^{t)}_\1 f^+  \cr
=&\ -~2 ( D^{(s}_\2 \ln f^+ +  D^{(s}_\1 f^0 ) D^{t)}_\1 F^- ,\quad
\hbox {using (17a),}\cr
=&\ -~2 D^{(s}_{{\dot 1}} F^0 ~ D^{t)}_{{\dot 1}} F^- .
\cr}$$ The second equation, on using the equations of motion for $f^0$ and
$F^-$ yield
$$\eqalign{
& D^{(s}_\2 D^{t)}_\1 F^0 \cr & =
  - D^{(s}_\1 f^-D^{t)}_\1  f^+
  - D^{(s}_\2 (F^0-f^0) f^+ D^{t)}_\1 F^-
  - (F^0-f^0) D^{(s}_\2 f^+ D^{t)}_\1 F^-     \cr & \hskip 3.5cm
  - (F^0-f^0) f^+ D^{(s}_\1 F^0 D^{t)}_\1 F^- ,\hbox{and using (17b)},\cr
& =\ - \left(  (f^+)^2 D^{(s}_\1 f^-
	      + D^{(s}_\2 (F^0 - f^0) f^+
	      + (F^0 - f^0) D^{(s}_\1 (F^0 - f^0) \right) D^{t)}_\1 F^- ,\cr
& =\ - D^{(s}_{{\dot 1}} F^+ ~ D^{t)}_{{\dot 1}} F^- , \hbox{using (17c)}
.\cr}$$
Finally (17c) implies $$\eqalign{
 D^{(s}_\2 D^{t)}_\1  F^+
 =&\  f^+ D^{(s}_\2 (F^0 - f^0) D^{t)}_\1 (F^0 - f^0)        \cr
  &\  +  D^{(s}_\2 f^+ \{ (F^0 - f^0) D^{t)}_\1 (F^0 - f^0)
                           + 2 f^+ D^{t)}_\1 f^-
			   + D^{t)}_\2 (F^0 - f^0) \}           \cr
  &\  +  (F^0 - f^0) f^+ D^{(s}_\2 D^{t)}_\1 (F^0 - f^0)
      + (f^+)^2 D^{(s}_\2 D^{t)}_\1  f^-   .\cr}$$
Using (17c) the first term on the right is equal to
$$ \{ D^{(s}_\1 F^+  - (f^+)^2 D^{(s}_\1 f^- \} D^{t)}_\1  (F^0 - f^0) ;$$
and using (17(b,c)) the second term may be written
$$ - \{ f^+ (D^{(s}_\1 f^0 + D^{(s}_\1 F^0 ) + (F^0 - f^0)D^{(s}_\1 f^+ \}
\{(f^+)^{-1} D^{t)}_\1 F^+ f^+ D^{t)}_\1 f^- \}.$$
Gathering all the terms and using the equations of motion for $f^0, f^-$
and $F^0$, we obtain $$
D^{(s}_\2 D^{t)}_\1  F^+ =\ 2 D^{(s}_{{\dot 1}} F^+\ D^{t)}_{{\dot 1}} F^0
.\eqno\square$$

An equivalent proof of invariance of equations (6) under the
transformation (17) is also implied by the following.
\vskip 5pt
{\it The lagrangian density whose variation yields equations (14b), }
$$ {\cal L}_{A_1}(f) = \partial_{\a \2}f^0 \partial^{\a}_{\1}f^0
			+  \partial_{\a \2}f^+ \partial^{\a}_{\1}f^-
			-  f^+  \partial_{\a\1}f^0  \partial^{\a}_\1 f^-
		        +  f^0  \partial_{\a\1}f^+  \partial^{\a}_\1 f^-
,$${\it equivalent to (7) up to a divergence, changes under the \Bt  (17)
by a total divergence, namely, }
$$ \eqalign{
& {\cal L}_{A_1}(F)  - {\cal L}_{A_1}(f) \cr
& =\ \partial_{\a \2} ( F^0 \p^{\a}_\1 (F^0 - f^0) + F^+ \p^{\a}_\1 F^-
                      + \half ln f^+ \p^{\a}_\1(F^0 - f^0)^2
		       - ln f^+  \partial^{\a}_{\2} (F^0 - f^0) )  \cr
&\ +  \partial_{\a \1} ( - F^+ F^- \p^{\a}_\1 F^0
                      -  \half f^0 \partial^\a_\1 (F^0 - f^0)^2
                      -  {ln f^+ \over 3}  \partial^\a_\1 (F^0 - f^0)^3
		      -  {ln f^+ \over 2}  \partial^{\a}_{\2} (F^0 - f^0)^2)
.\cr}$$
{\it Proof.}
Using the equations of motion (14b),
$$ {\cal L}_{A_1}(F) =\
- F^-  \partial_{\a\1}F^0  \partial^{\a}_\1 F^+  ,$$
up to divergence terms.  Now inserting (17a) and the even equations
equivalent to (17b,c)
$$\eqalign{
 \p_{\a\1} F^0 =&  - \p_{\a\1} f^0 \ -\ \p_{\a\2} \ln f^+
                  - (F^0 - f^0)\p_{\a\1} ln f^+  \cr
 \p_{\a\1}F^+ =&  f^+ \{ (F^0 - f^0)\p_{\a\1} (F^0 - f^0)
 +\   f^+  \p_{\a\1} f^- \    +\  \p_{\a\2}(F^0 - f^0) \} .\cr} $$
immediately yields the result. $\quad \quad \square$
\vskip 20pt
{\bf 4. Explicit integration of the $A_1$ theory}
\vskip 20pt
We now construct a hierarchy of explicit solutions
$ f_n = \{f^0_n,f^+_n,f^-_n\}$ to the $A_1$ theory by iterating the
transformation (17) starting from the solution $ f_0 = \{\tau, \a_0, 0\}$
corresponding to a restriction of the gauge algebra to a solvable subalgebra,
with $\tau , \a_0$ satisfying the system of equations
$$ \eqalign{
 D^{(s}_{{\dot 2}} D^{t)}_{{\dot 1}} \tau ~
=& 0
\cr
 D^{(s}_{{\dot 2}} D^{t)}_{{\dot 1}} \a_0 ~
+~2 D^{(s}_{{\dot 1}} \tau ~ D^{t)}_{{\dot 1}} \a_0 =& 0
.\cr}\eqno(18)$$
The general solution $(\tau , \a_0 )$ to this system clearly depends on
four arbitrary functions (two each for $\tau$ and $ \a_0$). A restricted
example of solution depending on two arbitrary functions is given by

$$ \eqalign{
 \tau =&
 \int d\l \ \rho(x_{\a \1}+\l x_{\a \2},  \bt_{i\1}+\l\bt_{i\1}, \l)  \cr
  \a_0
=& \int d\l\  \s(x_{\a \1}+\l x_{\a \2},  \bt_{i\1}+\l\bt_{i\1}, \l) ~
 e^{ \{ \int {d\l' \over \l- \l'}
        \rho(x_{\a \1}+\l' x_{\a \2},  \bt_{i\1}+\l'\bt_{i\1}, \l')\}}
,\cr}
\eqno(19) $$ where $\rho,\s$ are arbitrary functions.
Now, taking $f = f_0$, the B\"acklund transfromation (9) can be easily
integrated, using a superpotential $\a_2$ satisfying
$$ D^t_\1 \a_{2} =\   \a_1 D^t_\1 \tau + D^t_\2 \a_1  ,\eqno(20)$$
to yield a new solution $F = f_1$ with components
$$ \eqalign{
f^-_1 =&\  -{1\over \a_0}  \cr
f^0_1 =&\  \tau -{\a_1 \over \a_0} \cr
f^+_1 =&\   {\a_1^2 \over \a_0} - \a_2
\cr}.\eqno(21)$$
In turn, taking $f = f_1$, the B\"acklund transformation is yet again
integrable and the process can be iterated, the transformation being
integrable for all the $f = f_n$  produced by iteration from the above $f_0$.
The integrations use a succession of superpotentials  $\a_n$
satisfying the relations
$$ D^t_\1 \a_{n+1} =  \a_n D^t_\1 \tau + D^t_\2 \a_n  .\eqno(22)$$
For the case of the special initial solution (19), these equations have
solutions
  $$ \a_n = \int d\l~  \l^n ~
      \s(x_{\a \1}+\l x_{\a \2},  \bt_{i\1}+\l\bt_{i\1}, \l)~
    e^{ \int {d\l' \over \l- \l'}
           \rho(x_{\a \1}+\l' x_{\a \2},  \bt_{i\1}+\l'\bt_{i\1}, \l')}
.\eqno(23)$$
The thus obtained solutions may be represented compactly in terms of
the determinant of a symmetric recursive matrix $M^{n+1}$
$$ D_{n+1} \equiv \dt M^{n+1} \equiv
       \dt  \pmatrix{  \alpha_0&\alpha_1&\alpha_2 &\ldots&\a_{n} \cr
	               \alpha_1&\alpha_2&\alpha_3 &\ldots&\a_{n+1}   \cr
                       \alpha_2&\alpha_3&\alpha_4 &\ldots&\a_{n+2} \cr
		       \vdots &\vdots &\vdots &\ddots &\vdots \cr
                     \alpha_{n-1}&\alpha_{n}&\alpha_{n+1}&\ldots&\a_{2n-2}\cr
                     \alpha_{n}&\alpha_{n+1}&\alpha_{n+2}&\ldots&\a_{2n-1}
\cr}\eqno(24a)$$
and of one of its minors $$
 \dot D_n \equiv   \dt
\pmatrix{  \alpha_0&\alpha_1&\alpha_2 &\ldots&\a_{n-1} \cr
           \alpha_1&\alpha_2&\alpha_3 &\ldots&\a_{n}   \cr
           \alpha_2&\alpha_3&\alpha_4 &\ldots&\a_{n+1} \cr
       \vdots &\vdots &\vdots &\ddots &\vdots \cr
         \alpha_{n-2}&\alpha_{n-1}&\alpha_n &\ldots&\a_{2n-3}\cr
         \alpha_{n}&\alpha_{n+1}&\alpha_{n+2} &\ldots&\a_{2n-1}\cr}
,\eqno(24b)   $$
where the dot denotes a derivation which shifts the index on $\a$ by one,
i.e. $\dot f(\{ \a_n \}) \equiv \sum_l \a_{l+1}\der{\a_l} f(\{ \a_n \})$.
Evidently, $ \dot D_n \equiv  - \tilde M^{(n+1)}_{n,n+1} ,$
where $\tilde M$ denotes the matrix adjugate to $M$.
\vskip 5pt
{\it
The recursive B\"acklund transformation
$$\eqalignno{
 f_{n+1}^-  =&\  -{1\over f_n^+} &(25a)\cr
 D^s_\1 f_{n+1}^0 =&\  - D^s_\1 f_n^0 \ -\ D^s_\2 \ln f_n^+
                  - (f_{n+1}^0 - f_n^0)D^s_\1 ln f_n^+  &(25b)\cr
 D^s_\1 f_{n+1}^+ =&\  f_n^+ \{ (f_{n+1}^0 - f_n^0)D^t_\1(f_{n+1}^0 - f_n^0)
        +\  f_n^+  D^s_\1 f_n^- \  +\  D^t_\2(f_{n+1}^0 - f_n^0) \}
,&(25c)\cr} $$
on iteration of the initial solution
$\{f^0_0 = \tau , f^+_0 = \a_0 , f^-_0 =  0\}$
yields, on integration, the hierarchy of solutions
$$\eqalignno{
f_n^- =&\  (-1)^n  {D_{n-1} \over D_n} &(26a)\cr
f_n^0 =&\  \tau -  {\dot D_n \over D_n} &(26b)\cr
f_n^+ =&\  (-1)^n  {D_{n+1} \over D_n} ,&(26c)\cr }$$
where  $ D_0 = 1, D_{-1} = 0, \dot D_0 = 0 $  and for $n\ge 1,
D_n$ and $\dot D_n $  are the determinants (24). }
\vskip 5pt
{\it Proof.}
Inserting (26) in (25b), we obtain the equations
$$  D^s_\1  \left( {\dot D_{n+1} \over D_n} \right)
  + \left( {D_{n+1} \over D_n} \right)^2  D^s_\1
                                \left( {\dot D_n \over D_{n+1}} \right)
  -  D^s_\2  \left( {\dot D_{n+1} \over D_n} \right)  =\  0
,\eqno(27)  $$
which may be proven following the proof of an analogous equation in appendix
B of [2]. In virtue of (22), eqs (27) are equivalent to the algebraic
equations
$$\eqalign{ \dot D_n   \der{\a_l} D_{n+1}  - D_{n+1}  \der{\a_l} \dot D_n
  +&\ \dot  D_{n+1}  \der{\a_l} D_n  -  D_n   \der{\a_l}\dot D_{n+1} \cr
    +&\  D_n \der{\a_{l-1}} D_{n+1}  - D_{n+1} \der{\a_{l-1}} D_n   =\  0
.\cr}\eqno(28)  $$
For any $l$, (28) contains three sets of determinants of the form appearing
in the following identity for determinants due to Jacobi:
$$\eqalign{
&\dt A^{(n-1)}\  \dt \pmatrix{   a&  \a_i     &b      \cr
		              \b_i&   A^{(n-1)} & \g_i  \cr
		                 c&  \d_i     & d     \cr} \cr
&=\dt \pmatrix{   a&  \a_i    \cr
              \b_i&   A^{(n-1)}\cr}
    \dt \pmatrix{  A^{(n-1)} & \g_i  \cr
                   \d_i     & d     \cr}
 - \dt \pmatrix{ \a_i     &b      \cr
		  A^{(n-1)} & \g_i  \cr}
   \dt \pmatrix{ \b_i&   A^{(n-1)}  \cr
		    c&  \d_i      \cr},\cr} \eqno(29)$$
where $A^{(n-1)}$ is any $(n-1)-$dimensional matrix and $i=1,...,(n-1)$;
and (28) may easily be seen to be a consequence of this identity.
The solution (26c) follows from (27) and the
and the recursion relation
$$ D_{n+1} D_{n-1} = \ddot D_n D_n  - \dot D_n \dot D_n    ,\eqno(30)$$
which is a special case of the identity (29).
The right side of (25c), on insertion of (26) and using the Jacobi relation
(30) to eliminate $D_{n-1}$
yields
$$\eqalign{
&{D_{n+1}^2 \over D_n^2} \dot D_n \left(
{D^s_\1\dot D_n \over D_n} - {\dot D_n D^s_\1 D_n \over D_n^2}
- {D^s_\1\dot D_{n+1} \over D_{n+1}}
+ {\dot D_{n+1} D^s_\1 D_{n+1} \over D_{n+1}}  \right) \cr
&- {\dot D_{n+1} D_{n+1} \over D_n} \left( {D^s_\1\dot D_n \over D_n}
- {\dot D_n D^s_\1 D_n \over D_n^2} - {D^s_\1\dot D_{n+1} \over D_{n+1}}
+ {\dot D_{n+1} D^s_\1 D_{n+1} \over D_{n+1}}  \right)    \cr
&+ {D_{n+1}^2 \over D_n^2}  \left(D^s_\1 \ddot D^s_\1 D_n
- {D^s_\1 \dot D_{n}^2 \over D_n}
- {\ddot D_n D^s_\1 D_{n+1} \over D_{n+1}}
+ {(\dot D_n)^2 D^s_\1 D_{n+1} \over D_n D_{n+1}}
+ {(\dot D_n)^2  D^s_\1 D_{n} \over D_n^2} \right)        \cr
&{D_{n+1}^2 \over D_n} D^s_\2 \left( { \dot D_n \over D_n}
- {\dot D_{n+1} \over D_{n+1}} \right)
\cr},$$
which on using (27) together with its derivation
$$
\left( D_{n+1} D^s_\2 D_{n} - D_n D^s_\2 D_{n+1} \right)\dot{}
=      \ddot D_{n+1} D^s_\1 D_{n} - D_n D^s_\1 \ddot D_{n+1}
  -  D_{n+1} D^s_\1 \ddot D_{n} + \ddot D_n D^s_\1 D_{n+1}
$$
reduces to
$$
{ D^s_\1 D_n \over D_n^2} ( \ddot D_{n+1} D_{n+1}
                           - \dot D_{n+1} \dot D_{n+1} )
- { D_{n+1} D^s_\1 \ddot D_{n+1} \over D_n}
+ {D^s_\1(\dot D_{n+1}^2)  \over D_n}
- {\dot D_{n+1}^2 D^s_\1 D_{n+1} \over D_n D_{n+1}} $$
$$ = - D^s_\1  \left( {D_{n+2} \over D_{n+1}} \right)$$
in virtue of (30), proving (26c). $\quad \quad \square $
\vskip 20pt
{\bf 5. Remarks}
\vskip 20pt
\item{a)} The infinitesimal forms of our \Bts  correspond to discrete symmetry
transformations of the self-duality equations. In fact, there exists an
infinite dimensional discrete group of symmetry transformations.
\item{b)}
Our solutions are not the general solutions of equations (6) since
they do not depend on sufficiently many arbitrary functions; e.g. for the
$A_1$ gauge theories, they depend on only four arbitrary functions
(the general solutions for $\tau$ and $\a_0$ depending on two arbitrary
functions each) rather than six. It remains to be seen whether
the harmonic superspace formulation of super-self-duality recently presented
[5] can be used to obtain more general solutions. Our superfield $f$, in
harmonic space, is the non-analytic superfield prepotential $V^{--}$
discussed in [5];
and as shown there, it may be used not only to reconstuct the spinor and
vector potentials of the theory, but also to directly constuct the superfield
strengths.
\item{c)} Following Witten's idea [12]
of intermingling self-dual and anti-self-dual data to obtain the full
(non-self-dual) $N=3$ equations, (6) and (7) were used in [4]
to construct a lagrangian and conservation laws for the full $N=3$ theory.
We expect the constructions of the present paper to
be of relevance for the full N=3 theory.
\item{d)} Equations (6,7) have also recently appeared in relation
to N=2,4 superstring theories [11], whose degrees of freedom, it has been
argued, describe N=4 super self-dual gauge theories. The relevance of our
discrete symmetries and solution hierarchy for string physics remains to be
investigated.
\item{e)} Apart from the harmonic superspace construction mentioned above,
there is an alternative construction of solutions using an auxiliary space.
Consider the first order integro-differential equation in the two auxiliary
parameters  $\l$ and $g$:
 $$\dd{\P(\l,g)}{g}  = \int {d\l' \over \l - \l'}~~[ \P(\l)~ ,~ \P(\l') ]  $$
In terms of solutions $\P(\l,g)$ of this equation,
$$ f = \int^1_0 dg~ \int d\l ~ \P(\l,g)     $$ solves (6) [8].
\vskip 20pt
{\it Acknowledgements.}
We should like to thank V. Ogievetsky and M. Saveliev for helpful discussions
and encouragement; and I. Musatov for help with algebraic computing.
\vskip 20pt
{\bf References}
\item{1. } Brihaye, Y., Fairlie, D.B., Nuyts, J., Yates, R.:
J. Math. Phys. 19 (1978) 2528.
\item{2. } Corrigan, E., Fairlie, D.B., Goddard, P., Yates, R.:
Comm. Math. Phys. 58 (1978) 223.
\item {3. } Devchand, C.:
 Nucl. Phys. B238 (1984) 333.
\item {4. } Devchand, C.:
J.Math.Phys. 30 (1989) 2978.
\item{5. } Devchand, C., Ogievetsky, V.: Super self-duality as
analyticity in harmonic superspace. CERN-TH. 6653/92, hep-th 9209120,
to appear in Phys.Lett.B.(1992).
\item{6. } Leznov, A.N.:
 B\"acklund transformation of self-dual
Yang-Mills fields for an arbitrary semisimple gauge algebra.
Protvino preprint IHEP 91-145.
\item{7. } Leznov, A.N.: Theor.Math.Fiz. 73 (1987) 302.
\item{8. } Leznov, A.N.: Protvino preprint IHEP 86-188.
\item{9. } Pohlmeyer, K.: Comm.Math.Phys. 72 (1980) 37.
\item{10. } Semikhatov, A.: Phys.Lett. 120B (1983) 171;
 Volovich, I.: Phys.Lett. 123B (1983) 329.
\item{11. } Siegel, W.: Stony Brook preprint ITP-SB-92-31.
\item{12. } Witten, E.: Phys.Lett. 77B (1978) 394.
\item{13. } Yang, C.N.: Phys.Rev.Lett. 38 (1977) 1377.
\end